\documentstyle[aps,array,twocolumn]{revtex}
\input epsf.sty
\begin{document}

\draft

\title{
An extrapolation method for shell model calculations}
\author{ Takahiro Mizusaki$^{1}$ and Masatoshi Imada$^{2}$}
\address {$^1$Institute of Natural Sciences, Senshu University, Higashimita, 
Tama, Kawasaki, Kanagawa, 214-8580, Japan}
\address {$^2$Institute for Solid State Physics, University of Tokyo,  
Kashiwanoha, Kashiwa, 277-8581, Japan}

\maketitle

\begin{abstract}
We propose a new shell model method, combining the Lanczos digonalization
and extrapolation method.
This method can give accurate shell model energy from a series of shell 
model calculations with various truncation spaces, in a well-controlled manner.
Its feasibility is demonstrated by taking the $fp$ shell calculations.
\\
\end{abstract}


Quantum many-body phenomena frequently reveal novel and emerging nature
beyond a simple extrapolation from the single-particle picture, mean-field
predictions, and exact diagonalization of  Hamiltonian matrices by such as
the Lanczos method.  Solving such many-body problems has been a common
challenging issue in elementary particle physics, nuclear structure physics,
condensed matter physics and others.
In these decades, diverse numerical methods, such as quantum Monte Carlo, 
stochastic diagonalization/variation, density matrix renormalization group (DMRG),
were proposed.
Among them, as a method combining the Path-integral formalism and 
diagonalization/renormalization group,
in nuclear structure physics, the Quantum Monte Carlo 
diagonalization (QMCD) method \cite{qmcd1,qmcd3} was proposed and was developed. 
In condensed matter physics, the Path-integral renormalization group (PIRG) method\cite{imada1} 
was proposed.
Though both share common procedure, in the latter method, an extrapolation method 
is utilized. 
This extrapolation method is significant by itself and can be applied, independently 
of other parts of the PIRG method.  
In the present paper, we combine this extrapolation method with the Lanczos diagonalization 
and propose a new shell model method. 

The shell model is one of the fundamental frameworks for nuclear many-body problem.
In the conventional shell model calculation, Hamiltonian matrix elements of the effective shell model
interaction are calculated by a complete basis set in a given 
shell model space, and then, diagonalization is carried out by the
Lanczos method. 
Such Lanczos shell model calculations have a long history over
a half century and have extended its realm of 
the application due to the progress of computer technologies. 
Nowadays, part of full $fp$ shell problem can be solved,
for instance, the yrast states of $^{52}$Fe, the $M$ scheme dimension of which 
in the complete $fp$ shell model space is 109,954,620,
was studied\cite{Ur}, while there still exist many nuclei not yet
explored in the full $fp$ shell due to their huge shell model dimensions. 

For such nuclei, truncation scheme is utilized for 
reducing the shell model space.  
In  the $fp$ shell calculations, due to relatively large gap of spherical
single particle energy between the $f_{7/2}$ orbit and others,
we restrict the shell model space by specifying the number of excited 
nucleons across this shell gap, that is,
$\oplus_{s\le t}(f_{7/2})^{A-40-s}(r)^s$ where $r$ means the set 
of the $f_{5/2}$, $p_{3/2}$ and $p_{1/2}$ orbits and $t$ is the maximum 
number of nucleons allowed to be excited.
As an example, we consider $^{48}$Cr within $fp$ shell model space,
of which shell model calculation needs about 2 million dimension. 
This was one of the state-of-the-art large-scale shell model calculations
in mid 1990's \cite{cr-caurier}.
In Fig. 1 (a) and 1(b), the ground state energies of the truncated space are plotted 
as functions of the $t$ value and the corresponding $M$ scheme dimensions, respectively.
The definite behavior of the convergence is not seen in Fig. 1, because 
energy eigenvalue as a function of the $t$ or $M$ scheme dimension does not 
have a well-defined scaling property.
Therefore, in general, we cannot predict the ground state energy in
the entire shell model space by extrapolating shell model results 
within smaller truncated spaces unless the truncation calculations give 
well-converged results. 
Hence, it is crucial to find a useful extrapolation scheme to
extract the correct results from truncation calculations.

\begin{figure}[h]
\begin{picture}(200,220)
    \put(0,0){\epsfxsize 180pt \epsfbox{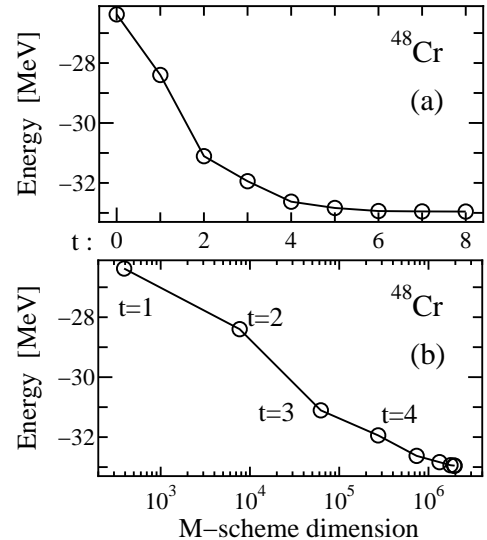}}
\end{picture}
\caption{Ground state energies of $^{48}$Cr as functions of
the $t$ values (a) and  the $M$ scheme dimensions (b). The labels $t$
are added for clarifying the truncation spaces in (b).
}
\end{figure}

In order to extrapolate the shell model results within smaller truncated spaces
into the complete shell model result,
a well-defined scaling property for energy eigenvalues is needed.
We define the difference $\delta E$ between the energy eigenvalue 
$\langle {\hat H} \rangle$ in a given truncated 
space and true energy eigenvalue $\langle {\hat H} \rangle _g$, that is,
$\delta E = \langle {\hat H} \rangle - \langle {\hat H} \rangle_g$.
The energy variance  $\Delta E$ in a given truncated spaces also defined as,
$\Delta E={{\left\langle {\hat H^2} \right\rangle -\left\langle {\hat H} \right\rangle ^2}
 \over {\left\langle {\hat H} \right\rangle ^2}}.$
The difference $\delta E$ vanishes linearly as a function of the energy variance $\Delta E$.
This relation was utilized in Refs. \cite{imada1} and \cite{sorella} and 
the summary of its proof can be seen
in Ref. \cite{imada2}. An approximate ground state $\left| \psi  \right\rangle $ can be
decomposed by the true eigenstate $\left| {\psi _g} \right\rangle$ and rest component 
$\left| {\psi _e} \right\rangle $  as
$\left| \psi  \right\rangle =c\left| {\psi _g} \right\rangle +d\left| {\psi _e} \right\rangle $
where $c^2+d^2=1$, and $ \left| {\psi _g} \right\rangle $ and $\left| {\psi _e} \right\rangle$
are orthnormalized states.
Up to $O(d^3)$, $\delta E\propto \Delta E$ is satisfied\cite{imada2}.
Therefore, this relation shows a good scaling if $\left| \psi  \right\rangle $ is a good
approximation of the true eigenstate.

\begin{figure}[h]
\begin{picture}(200,130)
    \put(0,0){\epsfxsize 180pt \epsfbox{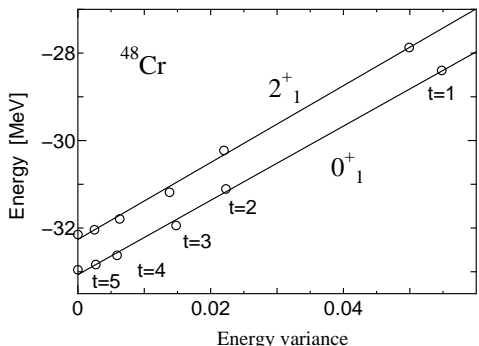}}
\end{picture}
\caption{Extrapolation of the energy to the zero energy variance for
the $0 ^{+}_{1}$ and $2^{+}_{1}$ states of $^{48}$Cr.
The $t$ values represent the truncation space.
}
\end{figure}

As an example, we reconsider the shell model results of $^{48}$Cr 
by the energy variance $\Delta E$ as the $x$-axis.
In Fig. 2, energy eigenvalues for various truncation spaces are plotted as a function of the
energy variance for the $0^+_1$ and $2^+_1$ states of $^{48}$Cr with the KB3 interaction\cite{kb3}. 
Linear relation between $\delta E$ and $\Delta E$ is seen from $t=1$.
By the results corresponding to the $t=1,2,3$, $t=3,4,5$, $t=4,5,6$,
 we can extrapolate the ground state energy as -33.148 $\pm$
0.15, -33.047 $\pm$ 0.02, -32.975 $\pm$ 0.02 MeV, respectively,
where these error-bars are given by the $\chi^2$ fitting.
As the exact ground state energy is -32.954 MeV, the extrapolation from $t=1,2,3$ results
is already a good value. Moreover, as the truncation space becomes larger, the error due to 
the extrapolation procedure becomes smaller. Thus the extrapolated energy is given with
error-bar, and its error can be well-controlled.
In the SMMC calculation\cite{smmc}, 
a different extrapolation method is utilized for 
overcoming the minus sign problem \cite{alhassid}. 
The extrapolated  energy of this nucleus is -32.3 $\pm$ 0.4 MeV\cite{cr48-smmc}.
This error comes from two reasons. One is a statistical uncertainties due to the Monte Carlo
sampling, the other is an systematic error due to the extrapolation  and finite temperature
corrections. 
Compared with the SMMC calculation, the present method has only a small systematic error.
Here, we comment that, extrapolated energies may become lower than the exact energy
because this process does not necessarily guarantee the variational principle. 

\begin{figure}[h]
\begin{picture}(200,280)
    \put(0,0){\epsfxsize 200pt \epsfbox{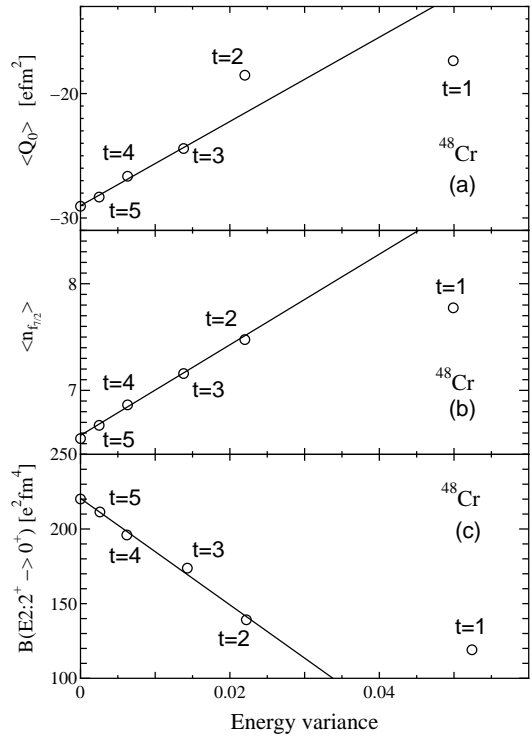}}
\end{picture}
\caption{Extrapolation of the quadrupole moment (a), occupation number
of the $f_{7/2}$ orbit (b) for the $2^{+}_{1}$ states 
and B(E2:$2^+_1 \rightarrow 0^+_1 )$ (c) 
to the zero energy variance.
The effective charges are taken as  $e_p=1.5e$ and $e_n=0.5e$.
}
\end{figure}

Next we consider the expectation value of various operators by the present
extrapolation method.
In Fig. 3, we show the quadrupole moment, occupation number and 
B(E2,$2^+_1 \rightarrow 0^+_1$) for  $^{48}$Cr.
As the first two values are expectation values by the same wavefunction, the 
$\Delta E$ can be uniquely defined. We can extrapolate them into zero energy variance
for quadrupole moments.
In Fig. 3(a), unlike energy, $t=1,2$ results seem to be rather poor. 
From the $t=1,2,3$, $t=3,4,5$ results,
the extrapolated quadrupole moments are -24.6 $\pm$ 4.1, -29.0 $\pm$ 0.3 $efm^2$, respectively.
The latter extrapolated value reaches almost exact one (-29.1 $efm^2$).
On the other hand, for the B(E2), the initial and final states are different.
Therefore, initial and final states have different $\Delta E$'s, which makes extrapolation 
procedure difficult.
In this case, however, the shell model spaces with the same $t$ value but 
different total magnetic quantum number have similar $\Delta E$.
Therefore, average $\Delta E$ works quite well in Fig. 3(c).
The extrapolated B(E2) from $t=2,3,4$ results is  220 $\pm$ 7 $e^2 fm^4$, which 
very well reproduces the exact one ( 220 $e^2fm^4$ ).

By this example, we conclude that energy vs. energy variance plot is useful to get 
an extrapolated energy from a series of shell model results with  various truncated spaces.
In addition to the energy estimates, the same extrapolation is also shown to be possible for
quadrupole moments, B(E2) and other quantities.

Moreover we comment on the present extrapolation from the structure of $^{48}$Cr.  
Low-lying yrast states of $^{48}$Cr is deformed\cite{cr-caurier}, and for its description,
the particle-hole excitation across the shell gap between the $f_{7/2}$ orbit and others 
is essential. 
Therefore it is interesting that the extrapolation from the truncated shell model results
with few particle-hole excitations works quite well.

\begin{figure}[h]
\begin{picture}(200,210)
    \put(0,0){\epsfxsize 200pt \epsfbox{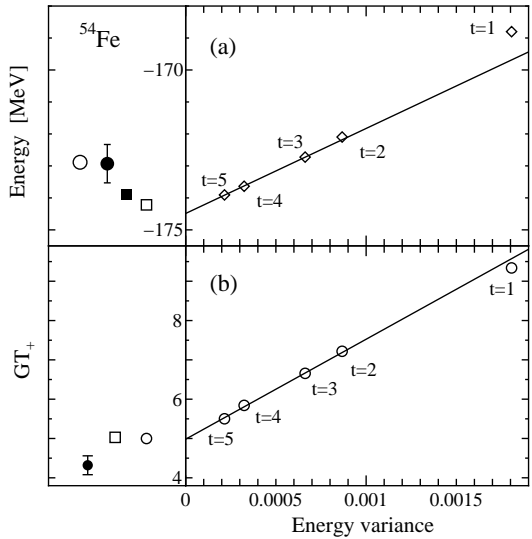}}
\end{picture}
\caption{(a) Extrapolation of the energy to the zero energy variance for 
$^{54}$Fe with other results of the SMMC\protect\cite{alhassid} (filled circle with error bar), 
statistical truncation\protect\cite{horoi}(open circle), $t=7$ truncated shell model 
(open square) and the QMCD\protect\cite{qmcd-fe} (filled square).
(b) Extrapolation of the total Gamow Teller strength  to the zero energy variance for 
$^{54}$Fe with other results of the extrapolation concerning to the $t$ value 
\protect\cite{gt-caurier} (open circle), $t=7$ truncated shell model (open square)
and the SMMC \protect\cite{alhassid} (filled circle with error bar). 
}
\end{figure}

We proceed further large-scale shell model  problem. The $M$ scheme dimension of $^{54}$Fe 
is about 0.35 billion and several recent methods\cite{alhassid,horoi} solved this shell model problem. 
For astrophysical interest, the Gamow Teller transition of this nucleus is known to be important.
The Gamow Teller transition also offers good benchmark test for the present extrapolation.
For the following calculation of  $^{54}$Fe, the FPD6 interaction\cite{fpd6} is taken 
as the effective interaction.
Fig. 4(a) shows the ground state energies of $t=$1,2,3,4 and 5 shell model spaces. 
The extrapolated ground state energy from $t=3,4,5$ is -174.49 $\pm$ 0.02  MeV.
Here we compare this value to the energies of other methods.
The ground state energy of the $t=7$ shell model spaces, of which $M$ scheme 
dimension is 91,848,462, is -174.217 MeV.
The extrapolated energy of the SMMC calculation \cite{alhassid} is -172.9 $\pm$ 0.6 MeV.
In  Ref.\cite{horoi}, from the statistical point of view, the $JT$ scheme bases
are selected and then the Hamiltonian matrix is diagonalized. This energy is about
-172.9 MeV \cite{horoi}. The ground state energy of the QMCD method with ten $J$-projected 
bases \cite{qmcd-fe} is -173.9 MeV. The present extrapolated energy is consistent with
the QMCD and $t=7$ truncated shell model energies.

In Fig. 4(b), the total GT+ strength is plotted. 
The extrapolated total GT+ from $t=3,4,5$ result is 4.98 $\pm$ 0.06.
The total GT+ strength of $t=7$ calculation is 5.03.
In Fig.2 of Ref. \cite{gt-caurier},  the total GT+ strength is plotted as 
a function of the $t$ value and it suggests that its total strength is about 5.
On the other hand, in Ref. \cite{alhassid}, the extrapolated value of the SMMC
calculation is 4.32 $\pm$ 0.24. Our extrapolated value is consistent with the former 
two values.
 
The $^{56}$Ni is a key nucleus for understanding the shell structure 
around the $fp$ shell region because it has a doubly magic structure, 
while its closed shell is suggested to be rather soft\cite{qmcd3,gsi}. 
Moreover, recently two deformed bands are discovered \cite{rudolph}.
This nucleus is quite interesting but the $M$ scheme dimension is quite 
large ( about $1.1 \times 10^9$ ). Then this nucleus
was a target of a state-of-the-art shell model calculation in the end of 
1990's \cite{qmcd3,gcmqmcd,rudolph}. 
Therefore it is also a good touchstone for the present extrapolation method.

\begin{figure}[h]
\begin{picture}(200,170)
    \put(0,0){\epsfxsize 220pt \epsfbox{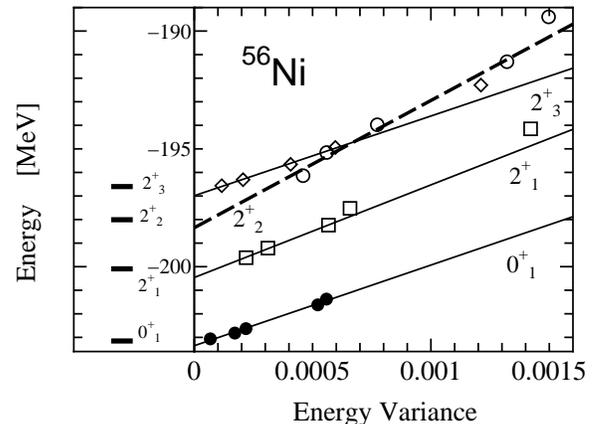}}
\end{picture}
\caption{Extrapolation of the energies of the $0^+_1$, $2^+_1$,
$2^+_2$ and $2^+_3$ states to the zero energy variance for $^{56}$Ni. 
For each state, the results of the $t=2 \sim 6$ truncated shell model spaces
are plotted. 
In the left part, the corresponding QMCD results\protect\cite{gcmqmcd} are shown.
}
\end{figure}

In Fig. 5, the energies of the $0^+_1$, $2^+_1$, $2^+_2$ and $2^+_3$ states 
with $t=2 \sim 6$ shell model spaces are plotted. 
In principle, the present extrapolation method can be applied to the 
excited states with the same quantum numbers without any difficulty.
Here we test its numerical feasibility. 
The extrapolated ground state energy from $t=4 \sim 6$ is -203.26 $\pm$ 0.06   MeV.
In Ref. \cite{qmcd-rev}, the ground state energy of the QMCD 
calculation is -203.152 MeV. 
The ground state energy of the $t=8$ truncated space is -203.178 MeV\cite{mshell-calc},
of which $M$ scheme dimension is 255,478,309. 
The extrapolated ground state energy is consistent with the results of the QMCD calculation
and the $t=8$ truncated shell model calculation.
The extrapolated $2^+_1$ energy is -200.46 $\pm$ 0.05 MeV from $t=4 \sim 6$ results, which is also 
consistent with the QMCD result\cite{gcmqmcd} and the $t=8$ truncated shell model result
(-200.075 MeV)\cite{mshell-calc}.

Next we investigate the excited $2^+$ energies. 
In the left part of the Fig. 5, the QMCD energies\cite{gcmqmcd} are shown.
These $2^+$ states are characterized by the quadrupole moments and the occupation numbers 
in the $f_{7/2}$ orbit.
The quadrupole moments are  21.6, -41.9 and 9.5 $efm^2$ for the  $2^+_1$, $2^+_2$ and $2^+_3$ 
states, respectively.
The occupation numbers are  12.9, 11.6 and 12.1. 
In Ref. \cite{gcmqmcd}, the $2^+_2$ state is concluded to be a  prolate deformed state. 
As the truncation scheme is based on the spherical single particle energies,
the deformed $2^+_2$ state is considered to be difficult to describe 
within small $t$ truncated spaces. 
For $t \le 6$, the states with negative quadrupole moment appear as the third $2^+$ state.
By plotting the shell model energies for the $t=2 \sim 6$ truncated spaces,
we can find the linear relations for $2^+_1$, $2^+_2$ and $2^+_3$ energies, respectively,
as shown in Fig. 5.
Therefore, for excited states, the present extrapolation method can be applied. 
Moreover Fig. 5 shows crossing between the $2^+_2$ and $2^+_3$ energies 
as a function of the energy variance, which means that, 
the present extrapolation method can correctly handle these excited states with different nature. 
The extrapolated energy of the $2^+_2$ state from $t=3 \sim 6$ results is 
-198.35 $\pm$ 0.35 MeV and extrapolated quadrupole moment and occupation number are 
-28.2 $\pm$ 8.5 $efm^2$ and 11.5 $\pm$ 0.6, respectively,  which are consistent with the QMCD values.
For more presice description of this prolate deformed state, $t=7$ results are needed, while
it is quite surprising that such a deformed excited state can be described
by the present extrapolation method based on the small truncated shell model spaces.

Finally we discuss a computational aspect. For the present extrapolation method,
in addition to the usual Lanczos procedure, the evaluation of 
the matrix element $\langle {\hat H}^2 \rangle$ is required.
To obtain this matrix element, we operate the Hamiltonian $H$ to the
$\left| \psi  \right\rangle$ of the $t$ truncated space in the ($t+2$) truncated space,
then we can evaluate the norm of the $ H \left| \psi  \right\rangle $. 
Therefore, the implementation of this extrapolation method is quite easy 
for a usual shell model program. 

In summary, we have proposed a new shell model method based on the Lanczos diagonalization 
and an extrapolation method. By taking the state-of-the-art $fp$ shell calculations
in the last decade, we have shown its numerical feasibility. 
The present extrapolation method can give accurate
energies, quadrupole moments and other quantities from a series of shell model
results with smaller truncated shell model spaces. 

The present extrapolation method was successfully applied to condensed matter
problems as a part of the PIRG method\cite{imada3}.
Therefore, it is intriguing to examine whether 
this extrapolation method is useful also in the QMCD method
for the nuclear shell model.
For instance, we expect more precise energy estimates 
from the QMCD calculations with smaller number of bases.  
Moreover, it might become quite useful in other methods of 
nuclear structure physics. Such researches are in progress.

We thank Professor  N. Yoshinaga for reading the manuscript and useful comments. 
This work was supported from 'Research for the Future Progam' by Japan Society 
for the Promotion of Science under grant RFTF97P01103, and was supported in part 
by Grant-in-Aid for Scientific Research (A)(2) (10304019), and one for Specially 
Promoted Research (13002001) from the Ministry of Education, Science and Culture.


\begin{references}
\bibitem{mshell-calc} These are the largest shell model calculations in our shell 
model code MSHELL\cite{mshell}.
\end{references}
\end{document}